\documentclass{raa_twocolumn}            

\usepackage{amsmath}	
\usepackage{amssymb}	
\usepackage[pagebackref=true]{hyperref}
\usepackage{pifont}
\usepackage{marvosym}
\usepackage{graphicx}
\usepackage{times}
\usepackage{txfonts}
\usepackage{natbib}
\usepackage{multirow}
\usepackage{enumitem}
\usepackage{url}
\usepackage{color}
\usepackage{xcolor}
\usepackage{orcidlink}
\bibpunct{(}{)}{;}{a}{}{,}
\hypersetup{
    colorlinks=true,
    linkcolor=blue,
    filecolor=blue,      
    urlcolor=blue,
    citecolor=blue
}
\defcitealias{kong2019}{K19}
\defcitealias{liu2019}{L19}
\defcitealias{li2021}{L21}
\defcitealias{xiang2022}{X22}

\graphicspath{{./figures}}

\begin{document}

\title{Measurements of the Diffuse Interstellar Bands at 5780, 5797, and 6614 \AA \ in the Hot Stellar Spectra of the LAMOST LRS DR10}

\volnopage{Vol.0 (20xx) No.0, 000--000}      
\setcounter{page}{1}          


\author{
Xiao-Xiao Ma \inst{1, 2}\orcidlink{0000-0002-9279-2783} \and
A-Li Luo \inst{1, 2}\orcidlink{0000-0001-7865-2648} \and
Jian-Jun Chen \inst{3, 2}\orcidlink{0000-0003-4525-1287} \and
Jing Chen \inst{4, 5}\orcidlink{0000-0001-8869-653X} \and
Jun-Chao Liang \inst{1, 2}\orcidlink{0009-0001-9085-8718}
}


\institute{
CAS Key Laboratory of Optical Astronomy, National Astronomical Observatories, Chinese Academy of Sciences, Beijing 100101, China \\
e-mail: \href{mailto:lal@nao.cas.cn}{lal@nao.cas.cn}
\and
School of Astronomy and Space Science, University of Chinese Academy of Sciences, Beijing 100049, China
\and
Key Laboratory of Space Astronomy and Technology, National Astronomical Observatories, Chinese Academy of Sciences, Beijing 100101, China \\
e-mail: \href{mailto:jjchen@nao.cas.cn}{jjchen@nao.cas.cn}
\and 
Nanjing Institute of Astronomical Optics \& Technology, Chinese Academy of Sciences, Nanjing 210042, China
\and
CAS Key Laboratory of Astronomical Optics \& Technology, Nanjing Institute of Astronomical Optics \& Technology, Nanjing 210042, China \\
\vs \no
   {\small Received 2025 Month Day; accepted 2025 Month Day}
}

\abstract{
Diffuse Interstellar Bands (DIBs) are crucial tracers of the interstellar medium (ISM), yet their carriers remain poorly understood. While large-scale surveys have advanced DIB studies in cool stellar spectra, measurements in hot stellar spectra are still limited. Using 287\,277 high signal-to-noise (S/N $>$ 50) hot stellar spectra from the tenth data release of the Large Sky Area Multi-Object Fiber Spectroscopic Telescope low-resolution spectroscopic survey (LAMOST LRS DR10), we systematically measured the three prominent optical DIBs at 5780, 5797, and 6614 \AA. We published three catalogs containing 285\,103, 279\,195, and 281\,146 valid measurements for the DIBs at 5780, 5797, and 6614 \AA, respectively. Among them, 112\,479, 25\,232, and 71\,048 are high-quality samples after rigorous quality control. To our knowledge, these are the largest hot-star DIB datasets in the northern sky. The catalogs provide spectral metadata, added astrometeric information, DIB profiles, and quality metrics. Our methodology and open-source pipeline ensure reproducibility, while the scale and precision of the data support future statistical studies. We anticipate that these catalogs will highlight the LAMOST's role in advancing DIB research and deepening our understanding of the ISM.
}

\authorrunning{Ma, X.-X. et al.}   
\titlerunning{DIB Measurements in LAMOST Hot Stellar Spectra}           
\maketitle

\section{Introduction}

Diffuse Interstellar Bands (DIBs) are a series of broad interstellar absorption features observed in spectra. To date, over 600 DIBs have been confirmed (\citealt{fan2019}; \citealt{hamano2022}), spanning wavelengths from the optical to the near-infrared (NIR). Investigating the carriers of these features is essential for understanding the chemical composition, evolution, and environment of the interstellar medium (\citealt{herbig1975}; \citealt{snow2006}; \citealt{fan2017}). Since the first discovery of the DIBs $\lambda\lambda$5780 and 5797 \footnote{DIBs are traditionally named after their approximate central wavelengths in air.} by \cite{heger1922} over a century ago, research has predominantly focused on macromolecular candidates such as polycyclic aromatic hydrocarbons (PAHs) and $\rm C_{60}^{+}$ (\citealt{tielens2008}; \citealt{cordiner2019}). Notably, $\rm C_{60}^{+}$ remains the only confirmed DIB carrier by \cite{campbell2015}, responsible for five NIR DIBs (refer to \citealt{linnartz2020} for a comprehensive review). 

However, laboratory challenges in synthesizing and characterizing such complex molecular candidates have hindered progress, prompting researchers to turn to astronomical observations to constrain potential carriers. For instance, \cite{ebenbichler2025} suggested that small molecules could be good DIB carrier candidates for the DIB $\lambda$6196 based on high-resolution hot star observations from the ESO diffuse interstellar bands large exploration survey (EDIBLES; \citealt{cox2017}). In addition to small-scale studies of DIBs using a few sample of high-resolution spectra of hot stars, recent advances in large-scale spectroscopic surveys (e.g., the Radial Velocity Experiment survey (RAVE; \citealt{steinmetz2006}), the Apache Point Observatory Galactic Evolution Experiment (APOGEE; \citealt{majewski2017}), and the Large Sky Area Multi-Object Fiber Spectroscopic Telescope (LAMOST; \citealt{cui2012}; \citealt{zhao2012}; \citealt{luo2012})), and improved DIB measurement techniques of cool stellar spectra (\citealt{yuan2012}; \citealt{kos2013}; \citealt{zhao2024}) have made it possible to conduct statistical studies of DIBs. For example, \cite{kos2014} presented the first pseudo-three-dimensional Galactic spatial distribution of the DIB $\lambda$8621 from the RAVE spectra of cool stars. \cite{zasowski2015} measured the DIB at 1.5273 $\mu$m in the APOGEE NIR spectra, revealing its kinematic properties for the first time. \cite{ma2024} established high-quality DIB measurement catalogs for the DIBs $\lambda\lambda$5780, 5797, and 6614 from over two million cool stellar spectra in the LAMOST low-resolution spectroscopic survey (LAMOST LRS), and systematically investigated their kinematics, sky distribution, and relationship with extinction.

Despite these advances, the DIB measurements in hot stellar spectra are limited or even absent in the aforementioned surveys. This is primarily due to the scarcity of hot stars or early-type stars in the stellar population, the lack of complete hot stellar spectral templates used to identify hot stars, and the source selection constraints of the surveys. For example, APOGEE focuses on red giants (\citealt{majewski2017}), while RAVE is mainly concentrated on late-type stars (\citealt{steinmetz2006}). Additionally, the DIB extractions from cool stellar spectra more or less suffer from residual stellar features (\citealt{zhao2024}). In contrast, hot stellar spectra exhibit fewer or even no stellar features (\citealt{gray2009}), which allows for direct and cleaner DIB measurements. With the LAMOST's efficient observational capabilities having amassed tens of millions of spectra, and with the significant progress in hot-star identification in LAMOST (\citealt{liu2019}; \citealt{li2021}; \citealt{xiang2022}; \citealt{kong2019}), a comprehensive survey of DIBs of hot stellar spectra in LAMOST has now become feasible. It provides an unprecedented opportunity to address this observational gap.

In this data paper, we presented the measurements of the three most prominent DIBs $\lambda\lambda$5780, 5797, and 6614 in the hot stellar spectra of the LAMOST LRS DR10 using the Gaussian profile fitting. Compared with the other DIBs in the LAMOST LRS wavelength coverage, the higher intensity of these three DIBs makes it easier to measure them (\citealt{fan2019}; \citealt{vogrincic2023}). We further published the data products of these DIB measurements online. The organization of this data paper is as follows: Section \ref{sec:data} introduces the data of the hot stellar spectra in the LAMOST LRS DR10, including the identification and selection of the hot stellar spectra. In section \ref{sec:measurement}, we outline the method of DIB measurement in the spectra of hot stars, including the data preprocessing and the fitting steps of DIB profiles. Section \ref{sec:data_products} elaborates on the released catalogs of the DIB measurements, with a focus on quality control, data formats, usage recommendations, and data characteristics. Section \ref{sec:conclusions} concludes the paper with a summary and future prospects. Finally, we provide the relevant links to the data and code in Section \ref{sec:data_availability}.

\section{Data} \label{sec:data}

LAMOST (\citealt{cui2012}; \citealt{luo2012}; \citealt{zhao2012}), located at the Xinglong Observatory in Hebei Province, China, is an adaptive reflecting Schmidt telescope with a four-meter aperture and a five-degree field of view. There are 4000 optical fibers uniformly equipped with 16 spectrometers on the same focal plane, which can theoretically acquire spectra of 4000 objects in a single exposure. The spectral resolution of LAMOST is classified into two modes for different scientific goals. One is the medium resolution (R $\approx$ 7500; \citealt{liu2020}). The other is the low resolution (R $\approx$ 1800; \citealt{deng2012}; \citealt{luo2015}), which spans the optical wavelength range from 3700 to 9000 \AA.

We utilized the spectra from the LAMOST LRS DR10 \footnote{\url{https://www.lamost.org/dr10}}. This data release publishes a total of 11\,817\,430 spectra, which encompasses 11\,473\,644 ($\sim$\,97.1\,\%) spectra of stars. So far, LAMOST has almost completed the spectroscopic survey of the entire northern sky, especially covering the Galactic plane where ISMs are dense. In this work, we focused on the hot stars in LAMOST due to the less contamination of stellar features in their spectra than the cool stars, which is beneficial for the DIB measurements. 


For the hot stars or early-type stars, namely, the O, B and A-type stars, although their stellar parameters are not available, the spectral classification, e.g., ``O'', ``B6'' and ``A3'', using the MK process is provided by the LAMOST stellar parameter pipeline (LASP; \citealt{luo2015}). Nonetheless, the number of the hot stars identified by the LASP is incomplete, because the spectral classification is not accurate enough for those hot stars, which can be further attributed to the scarcity of the standard template spectra of early-type stars. To address this issue, we re-compiled a catalog of the hot stars in LAMOST according to the official LAMOST catalog (hereafter LC), and several external ones by the teams that have searched for the hot stars in LAMOST through diverse methods. These works include \citealt{kong2019} (hereafter \citetalias{kong2019}), \citealt{liu2019} (hereafter \citetalias{liu2019}), \citealt{li2021} (hereafter \citetalias{li2021}), and \citealt{xiang2022} (hereafter \citetalias{xiang2022}). The comprehensive catalog is composed of two parts: the OB- and the A-type, which is established from the union of the catalogs mentioned above. Given that there may be some common spectra among these catalogs, we give priority to \citetalias{liu2019}, followed by \citetalias{li2021} and \citetalias{xiang2022}, and finally \citetalias{kong2019} and LC.

Table \ref{tab:hot} summarizes the statistics of these hot stellar spectra identified by various works. There are 680\,882 hot stellar spectra in total, including 105\,210 OB-type and 575\,672 A-type stellar spectra. We selected the spectra with the signal-to-noise ratio in $r$ band (S/N) greater than 50 as our targets to be measured. Such a selection is to ensure the reliability of the DIB measurements since the DIBs are weak features in the stellar spectra. Particularly, the depth is further degraded in the low-resolution spectra of LAMOST due to the instrumental broadening (\citealt{yuan2012}; \citealt{lan2015}; \citealt{ma2024}). After filtering, There are a total of 287\,277 hot stellar spectra that meet the S/N $> 50$.

\renewcommand{\arraystretch}{1.5}
\begin{table}
\bc
\caption{Statistics of the hot stellar spectra in LAMOST \label{tab:hot}}
\setlength\tabcolsep{10pt}
\begin{tabular}{crrrrrr}
\hline\hline
    & \multicolumn{2}{c}{The OB-type} & \multicolumn{2}{c}{The A-type} \\
From & $N_{\rm orig}$ & $N_{\rm sel}$  & $N_{\rm orig}$ & $N_{\rm sel}$ \\
\hline
\citetalias{liu2019}   \footnotemark[1] & 22\,901 & 21\,347 &     -    &     -    \\
\citetalias{li2021}    \footnotemark[2] &     208 &      11 &     -    &     -    \\
\citetalias{xiang2022} \footnotemark[3] & 92\,049 & 65\,657 & 362\,644 & 243\,024 \\
\citetalias{kong2019}  \footnotemark[4] & 10\,720 &  9\,786 &   6\,390 &      276 \\
            LC         \footnotemark[5] & 20\,170 &  8\,409 & 680\,989 & 332\,372 \\
\hline
\end{tabular}
\ec
\tablecomments{0.48\textwidth}{
For each catalog, $N_{\rm orig}$ indicates the original number of the hot stellar spectra, while $N_{\rm sel}$ refers to the number of the selected spectra. \\
\\
\footnotemark[1] A catalog of OB-type stars identified in the LAMOST LRS DR5. \\
\footnotemark[2] A catalog of Galactic O-type stars. The authors provided detailed information, including subclasses like ``O9.5 V''. \\
\footnotemark[3] A catalog of stellar labels for 330\,000 hot stellar stars from LAMOST LRS DR6. The authors determined various stellar parameters, including $T_{\rm eff}$. OB-type stars are those with $T_{\rm eff} > 10\,000$ K; otherwise, they are classified as A-type. \\
\footnotemark[4] A catalog of DB white dwarfs in LAMOST. The authors based it on LAMOST LRS DR5, but here an updated version from DR10 is used, available at \url{https://www.lamost.org/dr10/v1.0/catalogue}. If $T_{\rm eff} > 10\,000$ K (as estimated by the authors), the star is classified as OB-type; otherwise, it is A-type. \\
\footnotemark[5] The official general catalog of LAMOST. OB-type and A-type stars are selected based on the spectral classification provided by LASP. If the \texttt{subclass} keyword contains ``O'' or ``B'', the star is classified as OB-type. If it includes ``A'', the star is classified as A-type.
}
\end{table}

\section{Measurement} \label{sec:measurement}

Compared to cool stars, hot stellar spectra often exhibit fewer or even no stellar metal lines (\citealt{gray2009}), which prevents DIBs from blending with stellar features, thereby allowing for a forthright measurement of DIB profiles (\citealt{herbig1975}; \citealt{fan2019}; \citealt{smoker2023}). Following such a strategy, we directly measured the DIBs $\lambda \lambda$5780, 5797, and 6614 in the hot stellar spectra after the necessary preprocessing steps.

\begin{figure*}
\centering
\includegraphics[width=1.0\textwidth]{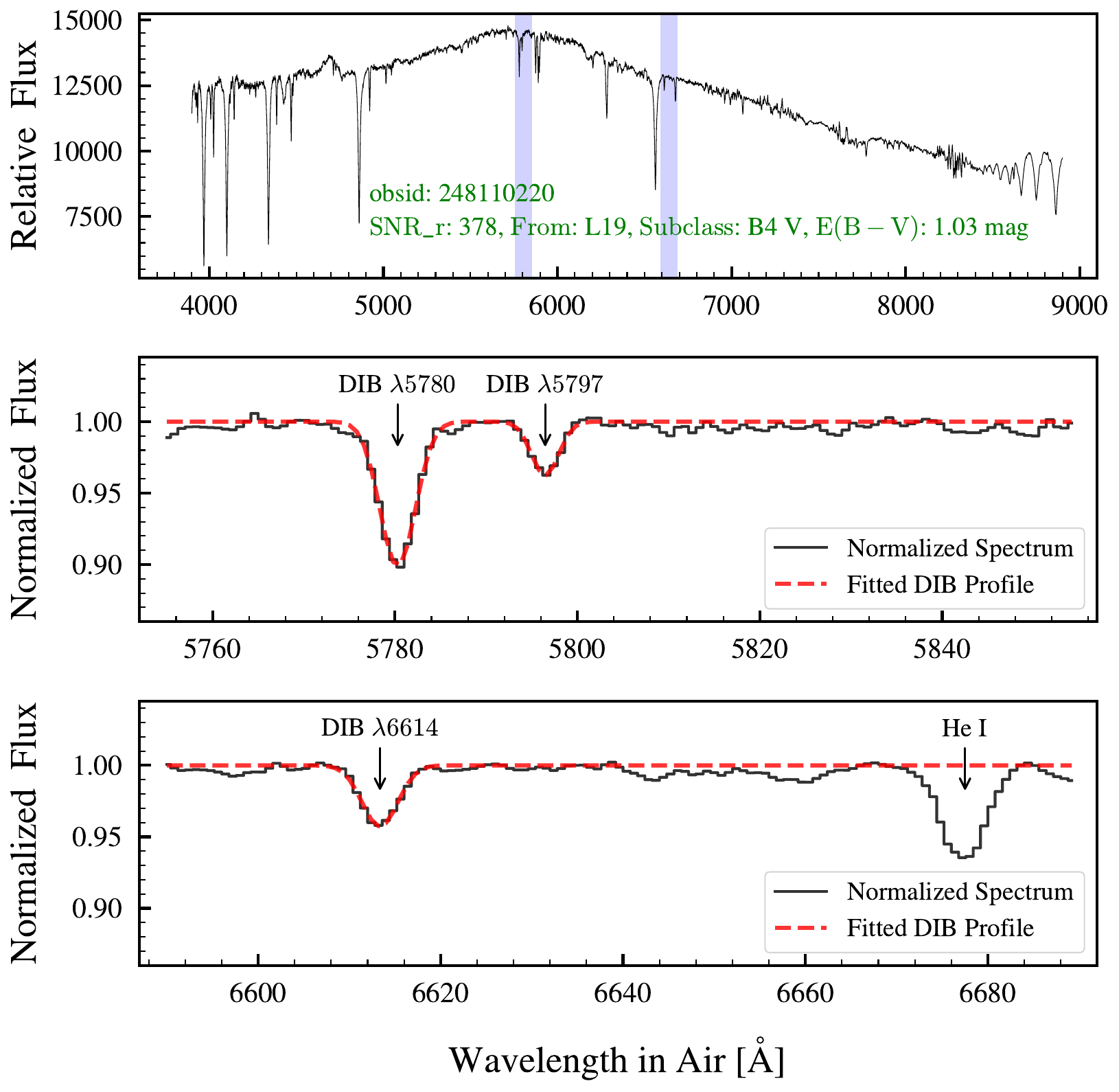}
\caption{
Instance of the DIB measurement in the HIS. The \textit{upper panel} shows a reddened spectrum with its unique ID in LAMOST (\texttt{obsid}) and extra information, i.e., the S/N in $r$ band, the source, the subdivision of spectral classification, and the extinction towards this star computed by the Python package \texttt{dustmaps} (\citealt{green2019}). In addition, the interesting DIB regions that cover the DIBs $\lambda \lambda$5780, 5797, and 6614 are also marked by the purple-shaded areas in the \textit{upper panel}. The \textit{middle panel} presents the spectrum (in black) containing the DIBs $\lambda \lambda$5780 and 5797 after the local continuum normalization, and its fitted Gaussian profiles (in red) by the MCMC sampling. The \textit{lower panel} is the same as the \textit{middle panel}, but for the DIB $\lambda$6614. The strong He\,\scalebox{0.8}{I} line at 6678 \AA \ which is usually seen in the B-type stellar spectra is also annotated in the \textit{lower panel}.
\label{fig:method}
}
\end{figure*}

We firstly converted the wavelength from vacuum to air. Then, we segmented the interesting DIB regions, namely, [5755, 5855] \AA \ for the DIBs $\lambda \lambda$5780 and 5797, and [6590, 6690] \AA \ for the DIB $\lambda$6614, as can be seen in the upper panel of Figure \ref{fig:method}. Each spectrum within the DIB region was rebinned to a step of 0.8 \AA, leading to 125 pixels for each DIB region. Subsequently, a fifth-order polynomial continuum normalization was applied to the rebinned spectrum. Such a fifth-order polynomial for continuum normalization is adopted as a practical and empirical (e.g., \citealt{ho2017}; \citealt{zhang2020,zhang2021}) choice that balances accuracy and efficiency across diverse spectral shapes, especially in large-scale automated DIB measurements. The error of the continuum-normalized flux was estimated by $\rm Err_{orig} / F_{cont}$ (\citealt{ho2017}; \citealt{zhang2020}), where $\rm Err_{orig}$ is the error of the original flux and $\rm F_{cont}$ is the polynomial-fitted continuum flux. The middle and the lower panels of Figure \ref{fig:method} illustrate the continuum-normalized spectra (in black) containing the DIBs $\lambda \lambda$5780, 5797, and 6614, respectively.

\begin{figure*}
\centering
\includegraphics[width=1.0\textwidth]{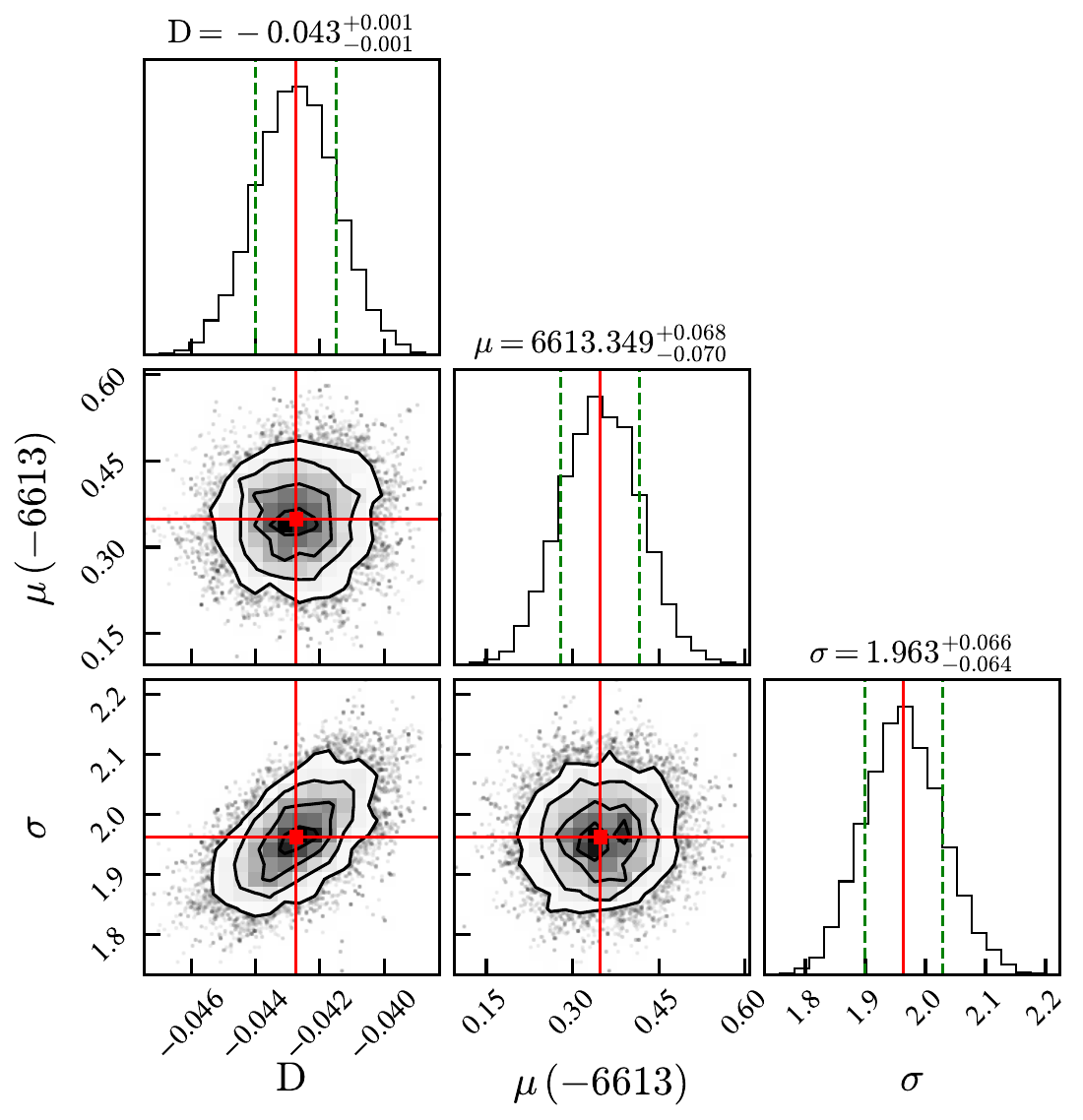}
\caption{
Corner plots of the MCMC sampling for the DIB $\lambda$6614 in the lower panel of Figure \ref{fig:method}. The posterior distributions of the depth, the central wavelength, and the Gaussian width are shown in the diagonal panels, while the two-dimensional histograms of the depth versus the central wavelength, the depth versus the Gaussian width, and the central wavelength versus the Gaussian width are displayed in the off-diagonal panels. The red solid lines in all the panels represent the median values of the posterior distributions. And the green dashed lines in the diagonal panels indicate the 16th and 84th percentiles, which represent the statistical uncertainties of the measurements.
\label{fig:mcmc}
}
\end{figure*}

After the aforementioned preprocessing steps, we performed the same DIB measurement of the continuum-normalized hot stellar spectra or hot interstellar spectra (HIS; \citealt{zhao2021a}) as that of the ISM residual spectra extracted from the cool stellar spectra or cool interstellar spectra (CIS), which has been elaborated in \citet{ma2024}. To be summed up, three consecutive steps guarantee the reliability of DIB measurement: (i) pre-detect whether the DIBs are present in the HIS; (ii) measure the DIBs using the curve fitting method; (iii) re-measure the DIBs based on the results of the curve fitting through the Markov chain Monte Carlo (MCMC) sampling. The middle and the lower panels of Figure \ref{fig:method} display the MCMC-fitted curves (in red) of the DIBs $\lambda\lambda$5780, 5797, and 6614 in the HIS, respectively. Figure \ref{fig:mcmc} illustrates the corresponding corner plots of the MCMC sampling for the DIB $\lambda$6614 in the lower panel of Figure \ref{fig:method}, which shows its posterior distributions of the depth, the central wavelength, and the Gaussian width.

For both the curve fitting and the MCMC sampling, the profile of DIBs in the HIS is characterized by a Gaussian function (\citealt{yuan2012}; \citealt{lan2015}; \citealt{farhang2019}) expressed as Equation \ref{eq:gaussian}.
\begin{equation} \label{eq:gaussian}
    f_{\theta}(x;\text{D},\mu,\sigma) = \text{D} \times \text{exp}\left(-\frac{(x-\mu)^2}{2\sigma^2}\right) + 1 \ ,
\end{equation}
where $x$ is the wavelength, and the parameters of DIB profiles, $\theta = \{\text{D}, \mu, \sigma\}$, mean the depth, the central wavelength, and the Gaussian width, respectively. The values of $\text{D}_{\text{pr}}$ and $\mu_{\text{pr}}$ derived from the step 1 (pre-detection), and the fixed value of $\sigma$ were used as the initial parameters for the step 2 (curve fitting). By similar token, the results of $\text{D}_{\text{cf}}$, $\mu_{\text{cf}}$, and $\sigma_{\text{cf}}$ figured out from the step 2 (curve fitting) were employed as the initial values of the step 3 (MCMC sampling). The median values $\text{D}_{\text{mc}}$, $\mu_{\text{mc}}$, and $\sigma_{\text{mc}}$ of the posterior distributions for depth, central wavelength, and Gaussian width, obtained through the MCMC estimation, were used as the final measurements of the DIBs, with the 16th and 84th percentiles representing their statistical uncertainties. The column ``Number of All'' of Table \ref{tab:stat} lists the summary of the entire valid DIB measurements in the HIS.

\renewcommand{\arraystretch}{1.5}
\begin{table}
\bc
\caption{Summary of the DIB Measurements in the HIS \label{tab:stat}}
\setlength{\tabcolsep}{8pt}
\begin{tabular}{crrr}
\hline\hline
DIB & Number of All & Number of HQ & HQ Ratio \\
\hline
$\lambda 5780$ & 285\,103 & 112\,479 & 39.5\% \\
$\lambda 5797$ & 279\,195 &  25\,232 &  9.0\% \\
$\lambda 6614$ & 281\,146 &  71\,048 & 25.3\% \\
\hline
\end{tabular}
\ec
\end{table}

\section{Data Products} \label{sec:data_products}

In this section, we present the released catalog of DIB measurements from the hot stellar spectra of the LAMOST LRS DR10. We begin with the quality control procedures used to identify the high-quality (HQ) DIB measurements. Next, we describe the data format and relevant metadata of the final released catalog in detail. We then provide the guidance and recommendations for the proper use of our data. Finally, we present the statistical characteristics of the data, including the spatial distribution of sightlines, histograms of extinction, distance, and equivalent width (EW), as well as correlations between DIB strengths and dust extinction, which serve as a validation of the measurements.

\subsection{Quality Control}  \label{subsec:qc}

To ensure consistency with the DIB measurement catalog obtained from CIS (\citealt{ma2024}) of LAMOST LRS, we applied the same selection criteria. The HQ DIB measurements are defined as those that meet the following criteria:
\begin{enumerate}
\item $\rm D_{mc} >  \frac{3}{S/N}$;
\item The coefficient of variation (CV) for the set $\rm \{D_{pr}, \ D_{cf}, \ D_{mc} \}$ is less than 10\%, with CV calculated as the standard deviation (STD) divided by the mean;
\item The STD of the set $\rm \{ \mu_{pr}, \ \mu_{cf}, \ \mu_{mc} \} < $ 0.5 \AA;
\item $\left| \rm \mu_{mc} - \lambda_0 \right| < $ 3 \AA, where $\rm \lambda_0$ represents the traditional central wavelengths of DIBs (\citealt{herbig1975}; \citealt{fan2019}; \citealt{vogrincic2023}), specifically $\rm \lambda_0 = 5780.6, \ 5797.1, \ 6613.6$ \AA \ for the DIBs $\lambda \lambda$5780, 5797, and 6614, respectively;
\item 1 $< \rm \sigma_{mc} < $ 3 \AA;
\item $\rm EW_{DIB} / EW_{cont} > $ 33 \% for the DIB $\lambda 5780$ and $>$ 25 \% for the DIBs $\lambda \lambda$5797 and 6614. Here, $\rm EW_{DIB}$ denotes the equivalent width of the DIB, computed as $\sqrt{2\pi} \times \sigma_{\text{mc}} \times \text{D}_{\text{mc}}$, and $\rm EW_{cont}$ represents the integrated equivalent width of the HIS outside the DIB region, defined as $\rm [\mu_{mc} - 3 \sigma_{mc}, \ \mu_{mc} + 3 \sigma_{mc}]$.
\end{enumerate}
After applying the above selection criteria, we obtained the HQ measurements of the DIBs $\lambda \lambda$5780, 5797, and 6614 in the HIS, with the exact numbers set out in Table \ref{tab:stat}. Figure \ref{fig:method} demonstrates an example of the high-quality DIB measurements obtained after applying the selection criteria described above.

Different from \cite{ma2024}, we defined only the HQ measurements and not further classified a marginal-quality (MQ) category. However, we released all the valid DIB measurements in the HIS  (the ``Number of All'' column in Table \ref{tab:stat}) and provided several metrics for quality assessment. This allows users to flexibly select data based on their specific scientific goals and accuracy requirements. For detailed descriptions of these metric columns, refer to Section \ref{subsec:data_info}, and for usage recommendations, see Section \ref{subsec:data_usage}.

\begin{figure*}
\centering
\includegraphics[width=1.0\textwidth]{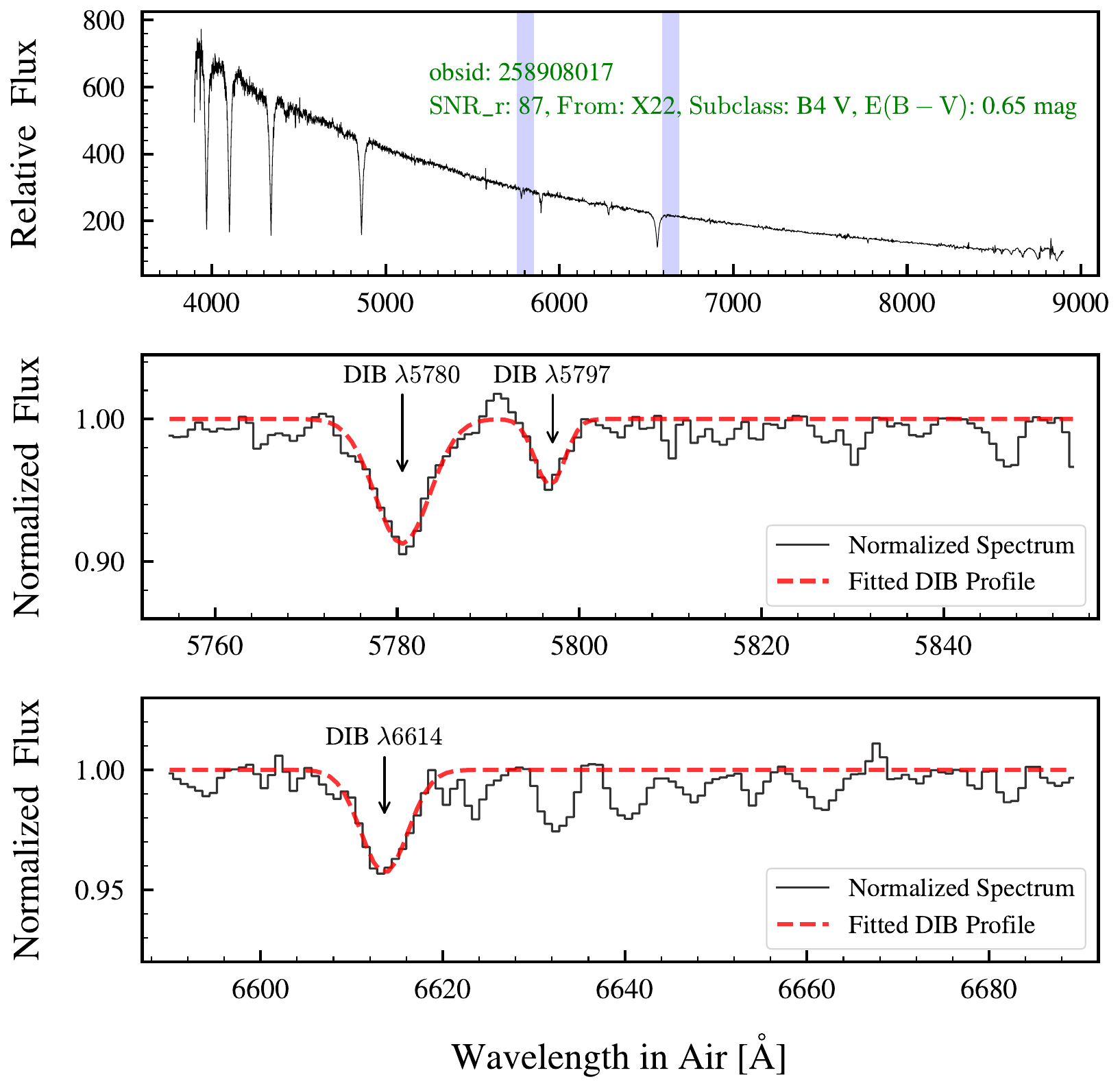}
\caption{
Same as Figure \ref{fig:method}, but showing an example of an intermediate-quality measurement with $\rm Err(EW_{DIB}) / EW_{DIB} < 1$.
\label{fig:intermediate_quality}
}
\end{figure*}

\begin{figure*}
\centering
\includegraphics[width=1.0\textwidth]{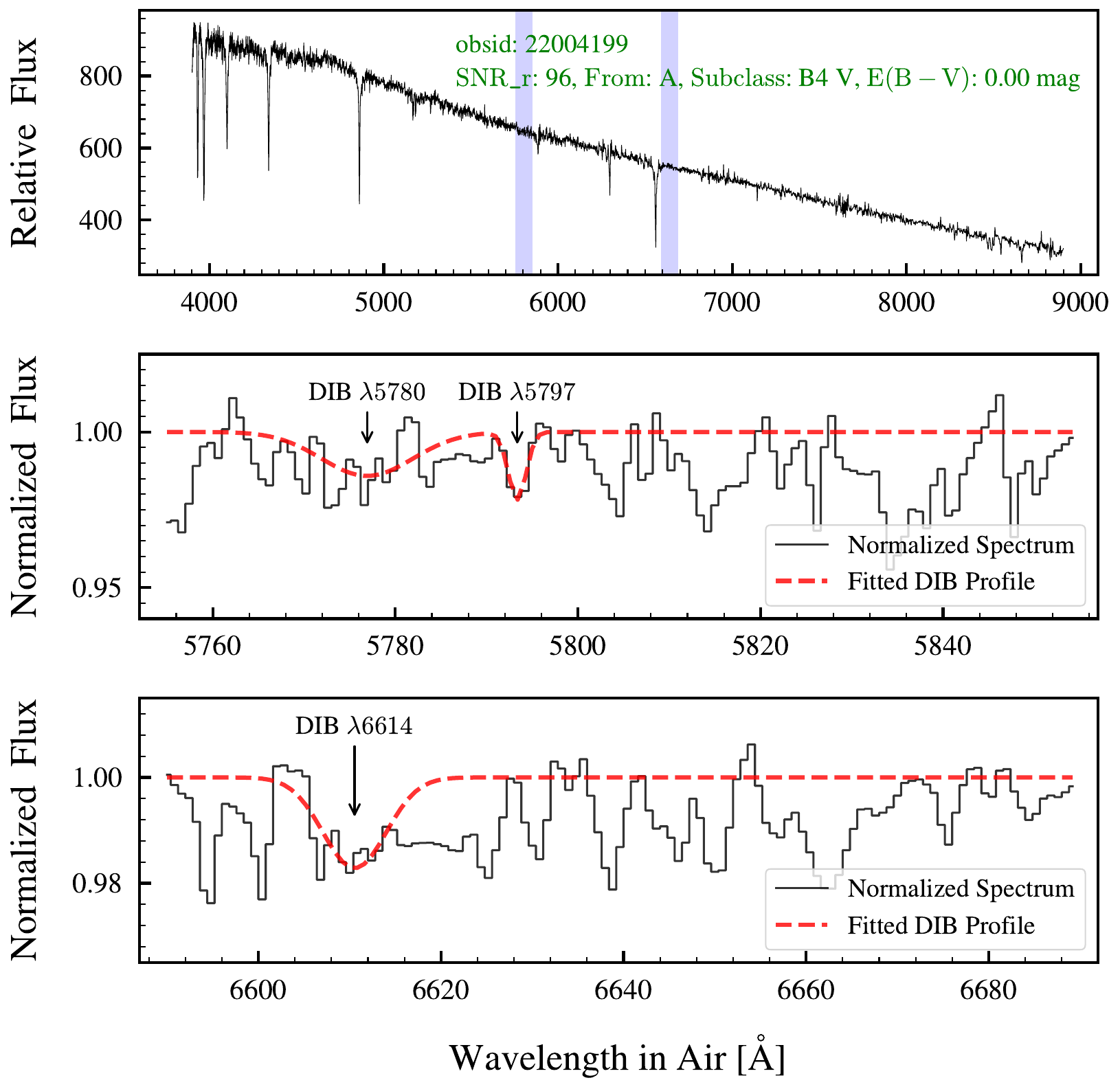}
\caption{
Same as Figure \ref{fig:method}, but showing an example of a low-quality measurement with $\rm Err(EW_{DIB}) / EW_{DIB} > 1$.
\label{fig:low_quality}
}
\end{figure*}

Although we did not perform a classification of samples beyond the high-quality group, we adopt a simple threshold of $\rm Err(EW_{DIB}) / EW_{DIB} = 1$ to provide users with an intuitive overview of the dataset. Based on this criterion, the measurements can be roughly divided into intermediate-quality ($\rm Err(EW_{DIB}) / EW_{DIB} < 1$) and low-quality ($\rm Err(EW_{DIB}) / EW_{DIB} > 1$) categories. Figures \ref{fig:intermediate_quality} and \ref{fig:low_quality} present examples of intermediate- and low-quality measurements, respectively, in the same format as Figure \ref{fig:method}. For intermediate-quality measurements, as shown in Figure \ref{fig:intermediate_quality}, although the DIB feature remains prominent, the reliability is reduced due to anomalies in the measured central wavelength and Gaussian width, as well as irregular absorption features outside the DIB region — likely caused by noise or stellar absorption lines. For low-quality measurements, as illustrated in Figure \ref{fig:low_quality}, the spectra are dominated by noise, and the DIB signal is either absent or undetectable. Even if a DIB profile can be fitted, its reliability is essentially negligible, and such cases should be discarded in practice. According to the threshold of $\rm Err(EW_{DIB}) / EW_{DIB} > 1$, less than 0.3\% low-quality samples of the entire dataset are identified. It is important to emphasize that this division serves illustrative purposes only and should not be regarded as a definitive quality classification scheme for specific scientific analyses. We have released additional quality assessment metrics (see Section \ref{subsec:data_info}), and users are encouraged to flexibly select and utilize the dataset according to their specific needs.

\subsection{Data Information} \label{subsec:data_info}

We released three catalogs containing the valid measurements for the DIBs $\lambda \lambda$5780, 5797, and 6614 in the HIS of the LAMOST LRS DR10, with their total counts listed in the ``Number of All'' column of Table \ref{tab:stat}. These tables share the same 22 columns, as detailed in Table \ref{tab:data}, which can be categorized into four groups:
\begin{enumerate}
\item Spectral metadata (Index 1-6), including identifier, celestial coordinate, and S/N;
\item Stellar properties (Index 7-9), such as spectral type, distance, and extinction;
\item DIB measurements (Index 10-17), estimated by the MCMC sampling in Section \ref{sec:measurement};
\item Quality assessment metrics (Index 18-22), which allow users to apply custom thresholds for data selection.
\end{enumerate}  

\renewcommand{\arraystretch}{1.5}
\begin{table*}
\bc
\caption{DIB Measurements in the HIS of the LAMOST LRS DR10 \label{tab:data}}
\setlength{\tabcolsep}{15pt}
\begin{tabular}{rcccl}
\hline\hline
Index & Label & Format & Units & Description \\
\hline
 1 & obsid            & Integer & -      & Unique spectrum ID in LAMOST \\
 2 & uid              & String  & -      & Unique source identifier in LAMOST \\
 3 & gaia\_source\_id & Integer & -      & Gaia DR3 source\_id \\
 4 & ra               & Float   & deg    & Right Ascension in J2000 \\
 5 & dec              & Float   & deg    & Declination in J2000 \\
 6 & snrr             & Float   & -      & S/N in $r$ band \\
 7 & class            & Integer & -      & Stellar class - 0 is OB type and 1 is A type \\
 8 & r\_med\_geo      & Float   & pc     & Geometric distance from \cite{bailer-jones2021} \\
 9 & ebv              & Float   & mag    & $\rm E(B-V)$ computed by 3D dust map (\citealt{green2019}) \\
10 & a                & Float   & -      & Depth of DIB \\
11 & aerr             & Float   & -      & Error of depth of DIB \\
12 & mu               & Float   & \AA    & Central wavelength of DIB \\
13 & muerr            & Float   & \AA    & Error of central wavelength of DIB \\
14 & sig              & Float   & \AA    & Gaussian width of DIB \\
15 & sigerr           & Float   & \AA    & Error of Gaussian width of DIB \\
16 & ew               & Float   & \AA    & Equivalent width of DIB \\
17 & ewerr            & Float   & \AA    & Error of equivalent width of DIB \\
18 & is\_hq           & Boolean & -      & Flag of high-quality DIB measurement \\
19 & int\_ew\_cont    & Float   & \AA    & Integrated equivalent width within the continuum region \\
20 & int\_ew\_dib     & Float   & \AA    & Integrated equivalent width within the DIB region \\
21 & cv\_a            & Float   & -      & Coeﬀicient of variation for the set $\rm \{D_{pr}, \ D_{cf}, \ D_{mc} \}$ \\
22 & std\_mu          & Float   & \AA    & Standard deviation of the set $\rm \{ \mu_{pr}, \ \mu_{cf}, \ \mu_{mc} \}$ \\
\hline
\end{tabular}
\ec
\end{table*}

\subsection{Data Usage} \label{subsec:data_usage}

Regarding data usage, we provide the following recommendations and considerations:  
\begin{enumerate}
\item The released catalogs are organized by spectra rather than by individual objects, meaning that the same object (grouped by the keyword \texttt{obsid}) may have multiple epoch-based DIB measurements. To reduce individual measurement uncertainties, we recommend using the mean or median of multiple measurements.
\item For studies that are not sensitive to sample volume, we suggest directly using the pre-selected HQ sample.
\item For statistical studies, such as the three-dimensional (3D) mapping or the relationship between DIB strength and extinction, users may relax the selection criteria based on the quality metrics (Index 19-22 in Table \ref{tab:data}) and the uncertainties in the DIB measurements to increase the sample size.
\item For the DIB $\lambda$5797 measurements, we strongly recommend using only the HQ sample. This is because DIB $\lambda$5797 is weaker in intensity compared to DIBs $\lambda\lambda$5780 and 6614 (\citealt{fan2019}), and is more susceptible to contamination from stellar spectral features, particularly the blending effect of a nearby Si\,\scalebox{0.8}{I} line at 5797.86 \AA \ (\citealt{herbig1975}).
\end{enumerate}

\subsection{Data Characteristics} \label{subsec:data_characteristics}

To provide an overview of the characteristics of the released catalog and demonstrate the reliability of our measurements, we present several visualizations of the data.

\begin{figure*}
\centering
\includegraphics[width=1.0\textwidth]{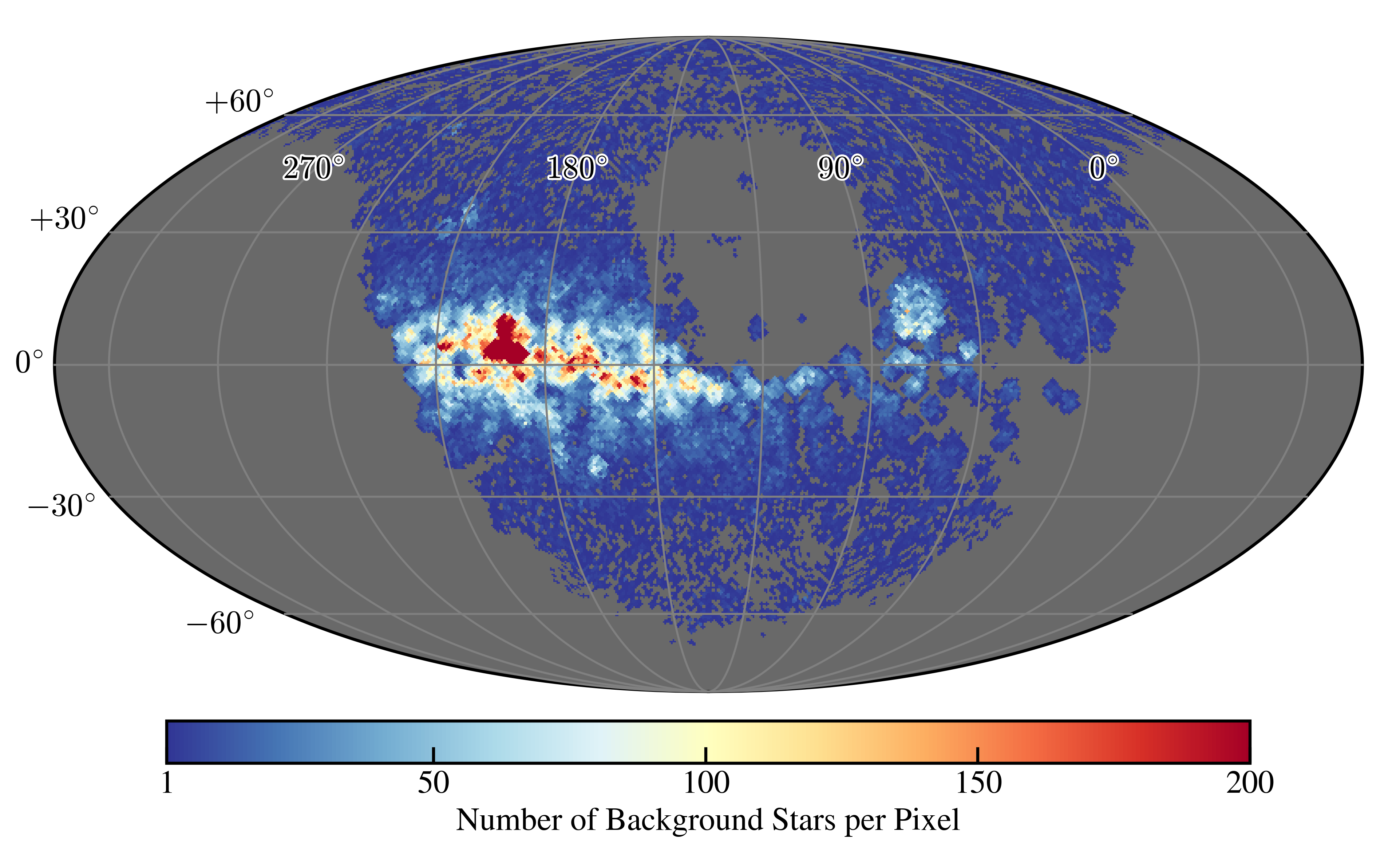}
\caption{
Galactic distribution of the sightlines. The colormap represents the number density of the background stars towards the sightlines. The resolution of the map is about 0.92$^{\circ}$ ($N_{\rm side}=64$) calculated by the HEALPix package. To highlight the overall distribution of the data, we truncated the number density within the 99th percentile.
\label{fig:spatial}
}
\end{figure*}

\begin{figure*}
\centering
\includegraphics[width=1.0\textwidth]{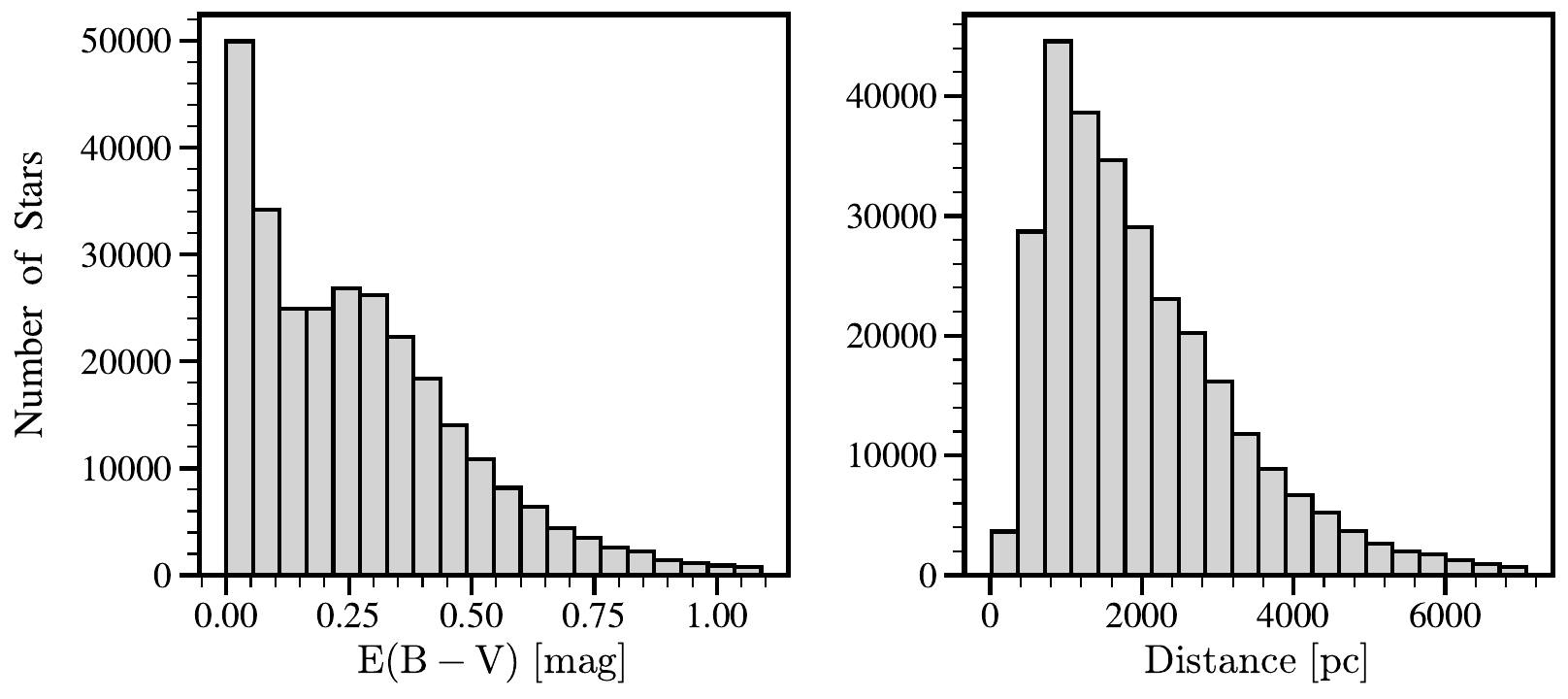}
\caption{
Histograms of extinction (left panel) and distance (right panel) for the background stars. To highlight the overall distribution of the data, we truncated the values within the 99th percentile.
\label{fig:hist_info}
}
\end{figure*}

We first present a statistical overview of the 287\,277 stellar spectra used for the DIB measurements. Figure \ref{fig:spatial} illustrates the Galactic distribution of these stars, constructed using the HEALPix (\citealt{gorski2005}). As shown, most of the background stars are concentrated on the Galactic plane, with a notable clustering in the Galactic anti-center. Figure \ref{fig:hist_info} further displays the extinction $\rm E(B-V)$ and distance distributions of these stars, revealing that the majority lie within the range of 1-3 kpc.

\begin{figure*}[htbp]
\centering
\includegraphics[width=1.0\textwidth]{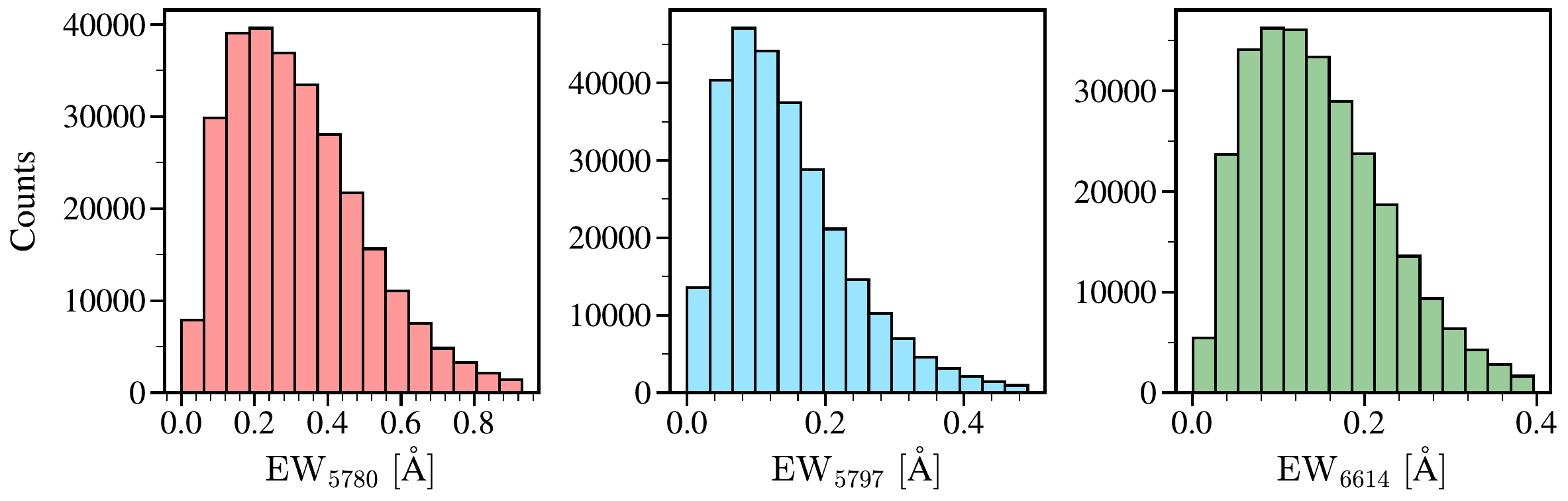}
\caption{
Histograms of EW for the DIBs $\lambda\lambda$5780 (left panel), 5797 (middle panel), and 6614 (right panel). To highlight the overall distribution of the data, we truncated the values within the 99th percentile.
\label{fig:hist_ew}
}
\end{figure*}

\begin{figure}[htbp]
\centering
\includegraphics[width=0.48\textwidth]{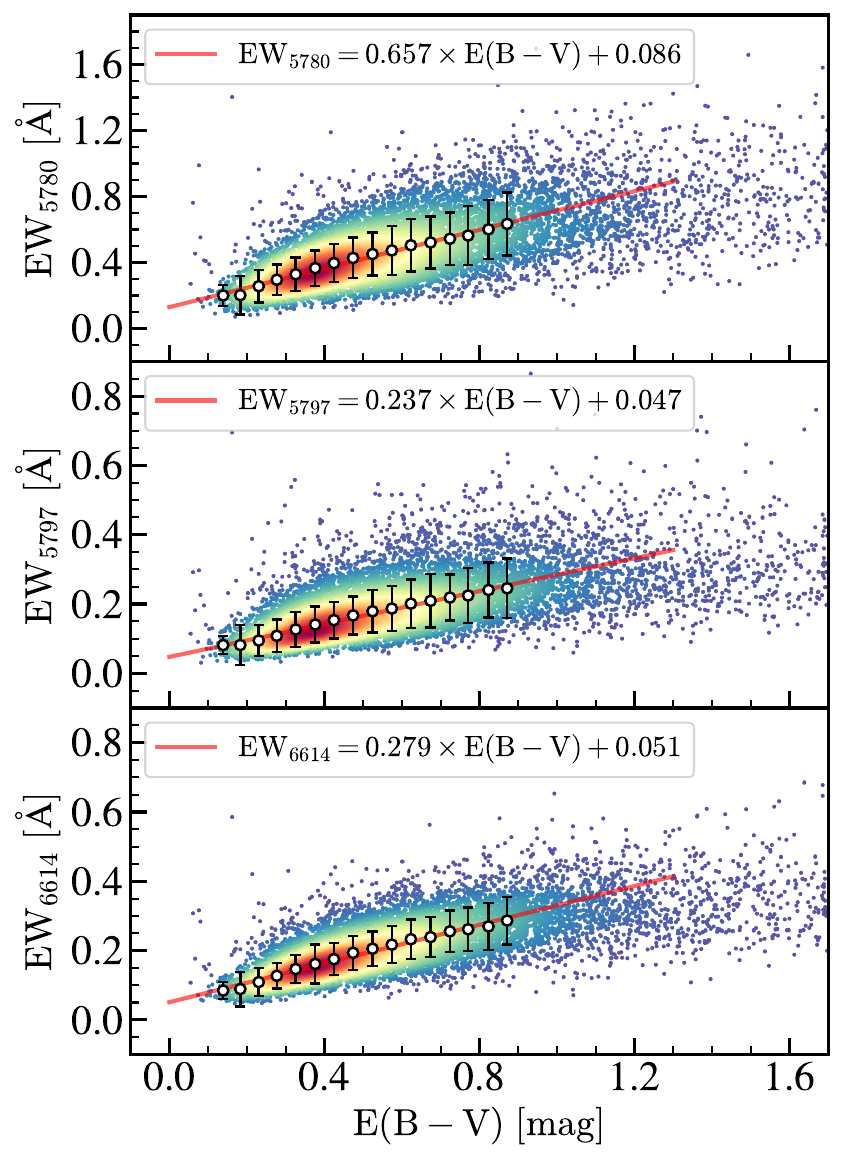}
\caption{
Gaussian KDE number density plot of the EWs of the DIBs $\lambda\lambda$5780, 5797, and 6614 (from top to bottom) as a function of the extinction $\rm E(B-V)$ estimated by \cite{green2019}. For each panel, the red solid line represents a linear fit to the median EW (which is marked with a white dot, and its standard deviation marked is with a black error bar) in each $\rm \Delta E(B-V) = 0.05 \ mag$ bin. The coefficients of the linear fit between the EW and $\rm E(B-V)$ are also annotated in each panel.
\label{fig:dib_ebv}
}
\end{figure}

\begin{figure}[htbp]
\centering
\includegraphics[width=0.48\textwidth]{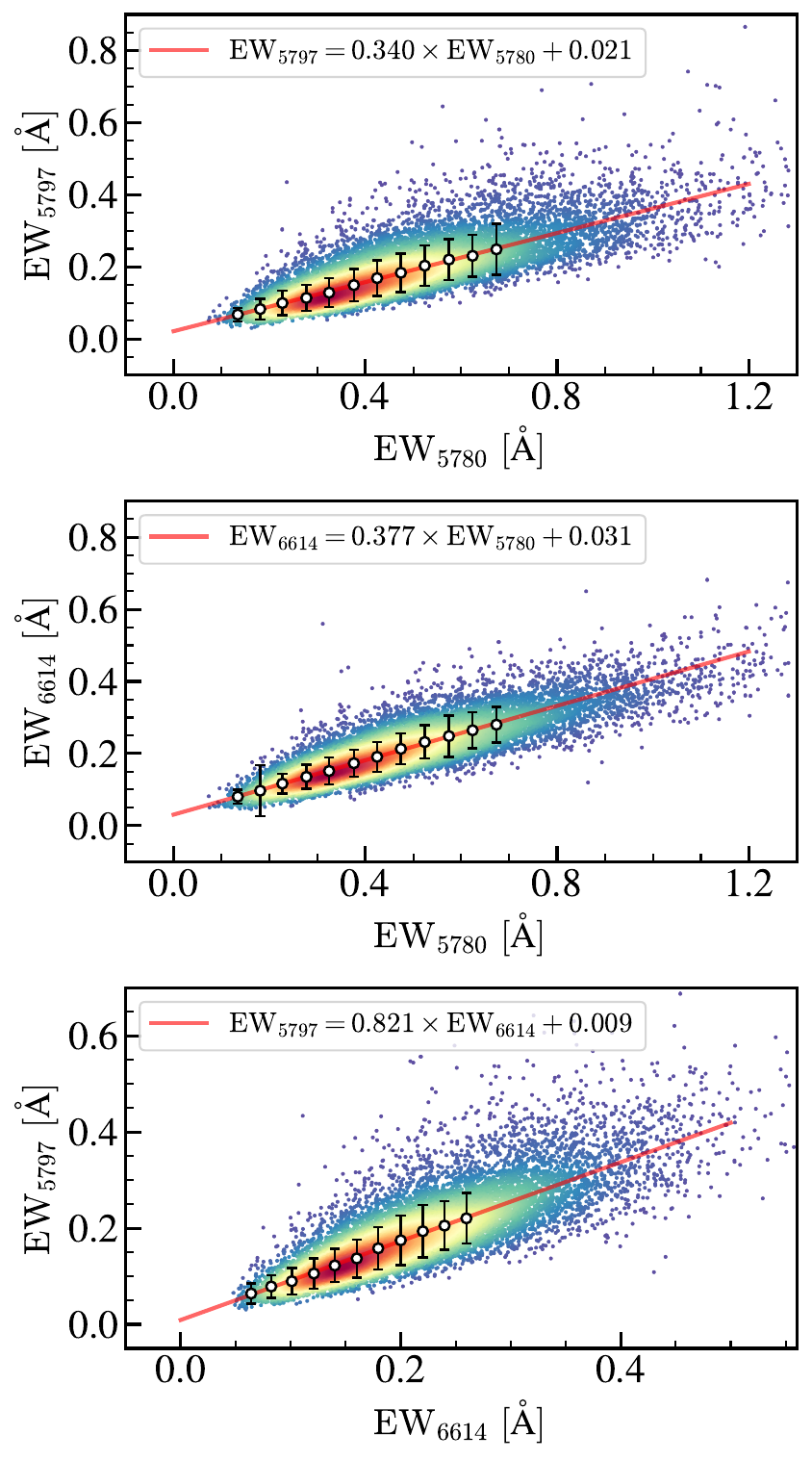}
\caption{
As in Fig. \ref{fig:dib_ebv}, but for the Gaussian KDE number density plot of EW versus EW. From the top panel to the bottom panel, the DIBs $\lambda\lambda$5780, 5797, and 6614 are compared with each other. For each panel, the red solid line represents a linear fit to the median EW in each $\rm \Delta EW = 0.05 \ \AA$ bin. Except for the bottom panel, the bin size of the EW is $\rm \Delta EW = 0.02 \ \AA$. The coefficients of the linear fit between the EWs are also annotated in each panel.
\label{fig:dib_dib}
}
\end{figure}

Subsequently, we provide a statistical summary of the DIB measurements. Figure \ref{fig:hist_ew} shows the histograms of the EWs for the DIBs $\lambda\lambda$5780, 5797, 6614. Using our HQ sample, we validated the relationships between EW and $\rm E(B-V)$, as well as among EWs of the different DIBs, as shown in Figures \ref{fig:dib_ebv} and \ref{fig:dib_dib}, respectively. All these correlations exhibit approximately linear trends. We also reported the coefficients from linear fitting, which are consistent with the results of \cite{ma2024}. However, we emphasize that Figures \ref{fig:dib_ebv} and \ref{fig:dib_dib} are intended primarily for illustrative purposes. As discussed in \cite{ma2024} and \cite{friedman2011}, simple linear correlations are insufficient to fully capture the underlying physical mechanisms. In our future work, we aim to employ more comprehensive analyses, such as 3D local density mapping and comparisons with other ISM tracers.

\section{Conclusions} \label{sec:conclusions}

In this data paper, we systematically measured the Gaussian profiles of the DIBs $\lambda \lambda$5780, 5797, and 6614 using the hot stellar spectra of the LAMOST LRS DR10. By carefully compiling a sample of 287\,277 hot stellar spectra with S/N $>$ 50 and applying a flowing measuring pipeline, we ensured reliable measurements of these DIBs. With the stringent quality control criteria, we identified 112\,479, 25\,232, and 71\,048 HQ measurements for the DIBs $\lambda \lambda$5780, 5797, and 6614, respectively.

This study extends and complements the work of \cite{ma2024} by measuring DIBs using LAMOST hot stellar spectra, whereas \cite{ma2024} focused on cool stellar spectra. Together, these efforts fully leverage the potential of LAMOST for interstellar medium studies. Furthermore, to the best of our knowledge, this dataset represents the largest sample of DIBs $\lambda \lambda$5780, 5797, and 6614 measurements in the northern sky, making it highly competitive in the field of DIB research. We encourage users to utilize the quality metrics to define their sample selection based on their specific research needs. By making full use of these data, users can explore and analyze various aspects of the interstellar medium, ultimately contributing to a better understanding of DIBs. Additionally, we welcome users to refine the measurement methods, as the code has been made publicly available on GitHub. Improving these methods can further enhance the accuracy of DIB measurements.

In the future, we will continue to refine and expand our work from both measurement and scientific perspectives. On the measurement side, as more LAMOST data become available, we will update the corresponding DIB measurements accordingly while also improving the measurement techniques. Additionally, we plan to incorporate more interstellar absorption features, such as DIB $\lambda$4430 and the Na\,\scalebox{0.8}{I} D doublet, to further enhance the dataset's applicability. On the scientific side, we aim to leverage the large-scale advantage of this dataset, particularly for statistical studies such as 3D local density mapping. We anticipate that this dataset will serve as a valuable resource for future research in DIB studies and beyond.

\section{Data Availability} \label{sec:data_availability}

The three catalogs of DIB measurements for the DIBs $\lambda \lambda$5780, 5797, and 6614 in the HIS of the LAMOST LRS DR10 are publicly available as FITS tables at \url{https://nadc.china-vo.org/res/r101573/}. The codes used for the DIB measurement pipeline are also provided at \url{https://github.com/iScottMark/LAMOST_DIB}.

The other data and the code used to generate the figures of this study are available from the corresponding author upon reasonable request.

\begin{acknowledgements}
This work is supported by National Key R\&D Program of China (Grant No. 2019YFA0405102) and National Natural Science Foundation of China (Grant Nos. 12273075 and 12090044). Guoshoujing Telescope (the Large Sky Area Multi-Object Fiber Spectroscopic Telescope, LAMOST) is a National Major Scientific Project built by the Chinese Academy of Sciences. Funding for the Project has been provided by the National Development and Reform Commission. LAMOST is operated and managed by the National Astronomical Observatories, Chinese Academy of Sciences.
\\
\\
\textit{Software}: Astropy (\citealt{2013A&A...558A..33A}; \citealt{2018AJ....156..123A}; \citealt{2022ApJ...935..167A}), VizieR \citep{2000A&AS..143...23O}, Simbad \citep{2000A&AS..143....9W}, TOPCAT \citep{2005ASPC..347...29T}, {\tt emcee} (\url{https://emcee.readthedocs.io/en/stable/}), {\tt uncertainties} (\url{https://uncertainties-python-package.readthedocs.io/en/latest/index.html})
\end{acknowledgements}

\clearpage
\bibliography{ms2025-0122}
\bibliographystyle{raa}

\end{document}